\documentclass[a4paper,11pt]{article}
\usepackage{jheppub}
\renewcommand\afterTocRuleSpace{\clearpage}
\usepackage{graphicx}
\usepackage{amsmath,amssymb}
\allowdisplaybreaks
\usepackage{booktabs}
\usepackage{placeins}
\usepackage{array}
\usepackage{url}
\usepackage{xcolor}
\usepackage{enumitem}
\newcommand{\Mc}{\mathcal{M}}
\newcommand{\Sn}{S_n}
\newcommand{\ip}[2]{\left(#1\,\middle|\,#2\right)}
\newcommand{\norm}[1]{\left\lVert #1\right\rVert}
\newcommand{\dpsi}{\delta\psi}

\newcommand{\pperp}{P_{\perp}}

\title{\boldmath Identifiability Limits of Gravitational Wave Phase Deviations: Multiclass Classification with a Multihead Neural Network}
\author[a]{Lavinia Heisenberg,}
\author[a]{Shayan Hemmatyar}
\affiliation[a]{Institute for Theoretical Physics, Heidelberg University,
Philosophenweg 16, 69120 Heidelberg, Germany}
\emailAdd{heisenberg@thphys.uni-heidelberg.de}
\emailAdd{hemmatyar@thphys.uni-heidelberg.de}

\abstract{%
In this work, we study how well gravitational wave phase deviations can be identified, using a neural network with two output heads for classification and regression. The classification head distinguishes general relativity (GR) from six parametrized post-Einsteinian phase families with exponents $b\in\{-7,-5,-3,-1,+1,+2\}$. The regression head predicts the logarithm of the coupling magnitude, $\log_{10}|\beta|$. The input to the neural network is a response function that quantifies the sensitivity of the waveform mismatch to waveform deformations. We use stationary Gaussian noise with the Advanced LIGO design spectrum for 262 sources and a measured Hanford spectrum from the third observing run (O3) for 132 sources. We also test injections into recorded Hanford strain. On the two Gaussian datasets, the network identifies the exponent of a deviation clearly above the noise with accuracies of $0.427$ and $0.329$. In order to interpret these results, we compare the network with an approximate classifier constructed from the predicted waveform residuals of the six phase families for the same sources. The network and this classifier often confuse the same families. By examining the residuals left by different phase corrections, we relate these errors to similarities between the underlying deviations. On selected recorded Hanford strain, networks trained on simulated or recorded noise give overall classification accuracies close to those obtained on simulated Hanford noise. Among thirteen neural network variants tested on the Advanced LIGO design dataset, the recurrent network achieves the highest mean classification accuracy for loud deviations, while all variants remain below the approximate classifier. The network reaches nearly the accuracy of the approximate classifier, which indicates that the identification is limited by the similarity of neighboring phase families rather than by the network, for known intrinsic source parameters.
}


\begin{document}
\maketitle

\section{Introduction}
\label{sec:intro}

Gravitational waves from inspiralling binary black holes allow us to test general relativity (GR) through the evolution of the waveform phase~\cite{will_testingalternativetheories_2014,abbott_observation_2016,abbott_tests_gw150914_2016,abbott_tests_2021}. A systematic description of generalizations of GR and their cosmological implications is given in~\cite{Heisenberg:2018vsk}. In parametrized tests of the inspiral, a departure from GR is described by additional terms in the waveform phase. The parametrized post-Einsteinian (ppE) framework~\cite{yunes_ppe_2009} organizes such terms by their frequency dependence, and a smooth taper confines the modification to the inspiral~\cite{mehta_fti_2023}. The coefficients can be constrained one at a time and, under the assumptions of a specified theory, related to its physical couplings~\cite{abbott_tests_2021,yunes_yagi_pretorius_2016,tahura_ppe_2018}.

The response function approach used here draws on earlier work in a cosmological context~\cite{heisenberg_simultaneouslysolving08tensionslatedarkenergy_2023,heisenberg_canlatetimeextensionssolve08tensions_2022}. This idea was developed for gravitational wave observables in~\cite{hemmatyar_paper1_2026}, where a response function was derived for the waveform mismatch. In this work, we use the mismatch response function as the input to a neural network with two outputs. The first output distinguishes GR from six ppE phase families, while the second regresses the logarithm of the magnitude of the coupling. The exponents $b\in\{-7,-5,-3,-1,+1,+2\}$ correspond to the post-Newtonian (PN) orders $-1$, $0$, $1$, $2$, $3$ and $3.5$.

We define a phase family by the ppE frequency dependence used to construct its correction. Corrections with the same exponent $b$ belong to the same family, while the coupling determines their strength. Different theories of gravity can predict the same leading frequency dependence. For example, two theories with leading ppE phase corrections proportional to $f^{-7/3}$ would both be represented by the $b=-7$ class, where $f$ is the gravitational wave frequency~\cite{tahura_ppe_2018}. The predicted class therefore indicates how the phase deviation varies with frequency. Identifying a specific theory requires additional information that distinguishes theories within that family, such as their predictions for further phase terms.

The choice of the GR template is an essential assumption of the analysis. We hold its intrinsic parameters at those of the injected source and determine its complex amplitude and arrival time from each noisy realization. The response function depends on source information that would have to be inferred in an observation. This construction is therefore an optimistic benchmark. A fit over intrinsic parameters could absorb additional components of the deviation, but the change in classification accuracy cannot be inferred from the residual norm alone. We do not establish an upper bound on the performance of an analysis that estimates those parameters from the data.

A low classification accuracy can reflect either the network or the similarity of the phase families themselves. In order to separate the two, we compare the network with a source-conditioned approximate reference constructed from the waveform residuals of the same phase families. We then study their local geometry, the confusion between classes, the detection of deviations with a fixed fraction of false positives, and the performance of the auxiliary regression head. We also inject the signals into recorded Hanford strain from the third observing run (O3) and compare several network variants. These additions address the noise model and the interpretation of the classification results, while retaining the stated restriction on the template parameters.

Related approaches address different parts of the problem. Narola et al.~\cite{narola_templatebgr_2022} enlarge a template search with modified phase coefficients. Wang et al.~\cite{wang_deeplearningbgr_2024} study the ability of networks trained on GR signals to detect signals with PN deviations, and Shahzad~\cite{shahzad_bgr_realnoise_2026} studies controlled waveform modulations in recorded LIGO noise. The neural post-Einsteinian framework of Xie et al.~\cite{xie_neuralppe_2024} uses a variational autoencoder to represent deviations in a continuous latent space and incorporates them into parameter estimation. Our analysis considers a discrete set of phase families using the mismatch response function evaluated with known intrinsic parameters. Its outputs are classification scores and an auxiliary point estimate, rather than posteriors from a joint inference of the source and deviation parameters.

The paper is organized as follows. Section~\ref{sec:framework} defines the signals, deviations, response function and sampling protocol. Section~\ref{sec:classifiers} describes the network, the reference and the evaluation metrics. We examine the geometry in section~\ref{sec:geometry} and present the Gaussian-noise results in section~\ref{sec:results}. Sections~\ref{sec:realnoise} and \ref{sec:variants} discuss recorded strain and network variants. We conclude in section~\ref{sec:discussion}. The derivation of the response function is given in appendix~\ref{app:response}.

\section{Framework}
\label{sec:framework}

We first specify the signals and the two noise constructions. We then introduce the six phase families and the response function used by both network heads.

\subsection{Signals, catalog and detectors}
\label{sec:signals}

We generate simulated gravitational wave signals from binary black hole systems~\cite{garcia-quiros_imrphenomxhm_2020,pratten_imrphenomxas_2020,pycbc_2024}. The masses are in the detector frame. The catalog provides the effective spin $\chi_{\rm eff}=(m_1\chi_1+m_2\chi_2)/(m_1+m_2)$ rather than the individual aligned components, and we set $\chi_1=\chi_2=\chi_{\rm eff}$. We use a single detector with the source face-on and directly overhead. The 262 sources are the binary black holes of the GWOSC catalog~\cite{gwosc} with published masses, effective spin and redshift and a secondary mass of at least $3\,M_\odot$, taken at the point estimates of GWTC-2.1, GWTC-3, GWTC-4 and GWTC-5.0~\cite{abbott_gwtc21_2024,abbott_gwtc3_2023,abbott_gwtc4_2025,lvk_gwtc5_2026}. We exclude GW190521 and GW200129\_065458, for which precession or eccentricity has been reported.

We consider the Advanced LIGO design sensitivity~\cite{aasi_advancedligo_2015} with a lower cutoff of $10$~Hz and a representative Hanford O3 sensitivity~\cite{ligo_t2000012} with a lower cutoff of $20$~Hz~\cite{abbott_gwtc2_2021,abbott_gwtc3_2023}. We refer to these as the aLIGO and Hanford rows. All 262 sources enter the aLIGO row, with optimal single-detector signal-to-noise ratios (SNRs) from $8.0$ to $136.6$. The Hanford row contains the 132 sources that satisfy the requirements of section~\ref{sec:protocol}, with SNRs from $8.0$ to $69.8$. In both rows, each source is injected into 100 independent realizations of stationary Gaussian noise drawn from the corresponding spectrum. Injections into recorded Hanford strain are described in section~\ref{sec:realnoise}.

\subsection{Deviations}
\label{sec:ppe}

We focus on the $(\ell,|m|)=(2,2)$ mode, since it gives the leading contribution to the gravitational radiation during the inspiral~\cite{garcia-quiros_imrphenomxhm_2020,pratten_imrphenomxas_2020}, and for this initial analysis we retain only this contribution in both the GR and the modified waveforms. In the stationary phase approximation, the GR waveform takes the form $h(f)=A(f)e^{i\psi(f)}$, with amplitude $A(f)$ and phase $\psi(f)$. The ppE framework~\cite{yunes_ppe_2009,tahura_ppe_2018} parametrizes deviations from GR as
\begin{equation}
h_{\rm ppE}(f)=h(f)\left[1+\alpha\,u^a\right]e^{i\beta u^b},\qquad u=(\pi\Mc f)^{1/3},
\label{eq:ppe_general}
\end{equation}
where $\Mc$ is the chirp mass in seconds. The exponents $a$ and $b$ determine the frequency dependence of the amplitude and phase corrections, while $\alpha$ and $\beta$ set their magnitudes, and GR is recovered for $\alpha=\beta=0$. Since the leading GR phase scales as $u^{-5}$, a phase correction with exponent $b$ enters at post-Newtonian order $(b+5)/2$~\cite{yunes_yagi_pretorius_2016}. Matched filtering is more sensitive to corrections of the phase than of the amplitude~\cite{tahura_ppe_2018,mehta_fti_2023}, and we therefore consider phase deviations only, $\alpha=0$. In addition, we confine the correction to the inspiral by replacing $u^b$ with a tapered profile $\phi_b(f)$, described below. The modified waveform is then $h(f)\exp[i\beta\phi_b(f)]$, which corresponds to
\begin{equation}
\psi(f)\longrightarrow\psi(f)+\dpsi(f),\qquad
\dpsi(f)=\beta\,\phi_b(f).
\label{eq:ppe}
\end{equation}
We refer to the real dimensionless coefficient $\beta$ as the coupling. Its relation to the physical couplings of a specific theory depends on that theory~\cite{yunes_yagi_pretorius_2016,tahura_ppe_2018}. Table~\ref{tab:families} lists the six exponents with their post-Newtonian orders, the form of $\phi_b$ during the inspiral, and examples of theories whose phase correction has that exponent.

\begin{table}[t]
\centering\small
\begin{tabular}{ccc>{\raggedright\arraybackslash}p{6.4cm}}
\toprule
$b$ & PN order & $\phi_b$ for $f\ll f_{\rm tape}$ & examples of theories \\
\midrule
$-7$ & $-1$ & $(\pi\Mc f)^{-7/3}$ & scalar-tensor, EdGB, Einstein-\ae ther, khronometric~\cite{yunes_yagi_pretorius_2016,tahura_ppe_2018} \\
$-5$ & $0$ & $(\pi\Mc f)^{-5/3}$ & Einstein-\ae ther, khronometric~\cite{yunes_yagi_pretorius_2016} \\
$-3$ & $1$ & $(\pi\Mc f)^{-1}$ & massive graviton~\cite{yunes_yagi_pretorius_2016} \\
$-1$ & $2$ & $(\pi\Mc f)^{-1/3}$ & dynamical Chern-Simons, noncommutative~\cite{yunes_yagi_pretorius_2016,tahura_ppe_2018} \\
$+1$ & $3$ & $(\pi\Mc f)^{1/3}$ & generic coefficient of the parametrized tests~\cite{abbott_tests_2021} \\
$+2$ & $3.5$ & $(\pi\Mc f)^{2/3}$ & generic coefficient of the parametrized tests~\cite{abbott_tests_2021} \\
\bottomrule
\end{tabular}
\caption{The six phase families. Below the taper, $\phi_b$ equals $u^b=(\pi\Mc f)^{b/3}$ up to a constant and a term linear in $f$. The last column gives examples of theories whose phase correction has the listed exponent. Several theories share an exponent, and Einstein-\ae ther and khronometric gravity contribute at two orders. EdGB denotes Einstein-dilaton Gauss-Bonnet gravity.}
\label{tab:families}
\end{table}

A power of $u$ describes the phase only during the inspiral, where the post-Newtonian expansion holds. Continued through merger and ringdown, where the waveform model is calibrated to numerical relativity, it would modify a regime in which the parametrization has no meaning. The parametrized tests of the LIGO--Virgo--KAGRA collaboration therefore switch the correction off before merger~\cite{abbott_tests_gwtc2_2021,abbott_tests_2021}, and we follow for this purpose the tapering prescription of~\cite{mehta_fti_2023}. It requires that the early-inspiral phase carries the full correction, that the post-inspiral phase reproduces the GR waveform up to a constant, and that the modified waveform is $C^2$ smooth at all frequencies.

The switch cannot act on the phase itself. A term $c_0+c_1f$ added to the phase is a shift of the coalescence phase and time and is not observable, so the physical content of a phase correction resides in its second frequency derivative, which alters the rate at which the frequency sweeps through the band. Multiplying $u^b$ by a window would introduce, through the derivatives of the window, a spurious contribution to this second derivative whose shape is set by the window rather than by the deviation. Instead, the window is applied to the second derivative, and the phase is recovered by integrating twice. In our convention,
\begin{equation}
\phi_b''(f)=W(f)\frac{d^2u^b}{df^2},\qquad
\phi_b(F)=\phi_b'(F)=0,
\label{eq:fti_d2}
\end{equation}
where primes denote frequency derivatives and $F$ is the upper end of the waveform generation grid. Thus,
\begin{equation}
\phi_b(f)=\int_f^F df'\int_{f'}^F df''\,W(f'')\frac{d^2u(f'')^b}{df''^2}
=\int_f^F (f'-f)\,W(f')\frac{d^2u(f')^b}{df'^2}\,df',
\label{eq:fti_integral}
\end{equation}
where the second form follows from interchanging the order of integration. The two integrations reconstruct the phase whose curvature is the windowed curvature of the ppE term. Below the window, where $W=1$, $\phi_b$ coincides with $u^b$ up to an affine function. Above it, where $W=0$, $\phi_b$ vanishes identically, and the transition between the two regimes is smooth. We fix the integration constants at the top of the grid, so that the correction vanishes above the taper. Any other choice differs from this one by an affine function, which the amplitude, phase and time match of every sample absorbs. For $b=3$, the power $u^3$ is linear in $f$ and its second derivative vanishes, so that $\phi_3=0$. This is expected, since a phase correction linear in frequency only shifts the arrival time, and it is the reason why $b=3$ is not among our exponents.

The window is
\begin{equation}
W(f)=\left[1+\exp\left(\frac{v-v_{\rm tape}}{\Delta v}\right)\right]^{-1},\qquad
\Delta v=\frac{128\eta}{3}\pi v_{\rm tape}^{6}\gamma N_{\rm GW},
\label{eq:window}
\end{equation}
where $v=(\pi Mf)^{1/3}$, $M$ is the detector-frame total mass in seconds, $\eta=m_1m_2/M^2$, and $v_{\rm tape}$ is the value of $v$ at $f_{\rm tape}$. It equals one well below $f_{\rm tape}$ and zero well above it, and its width follows from the requirement that the transition takes $N_{\rm GW}$ gravitational-wave cycles. At leading post-Newtonian order the number of cycles between two velocities is $N_{\rm GW}=(v_1^{-5}-v_2^{-5})/(32\pi\eta)$, which gives the expression for $\Delta v$ with $\gamma=3/20$. Since the taper acts close to merger, where this estimate is not reliable, $\gamma$ is instead treated as a phenomenological factor and set to $1/50$, and the outcome of the test is insensitive to $N_{\rm GW}$ between $0.8$ and $3$~\cite{mehta_fti_2023}. We adopt $f_{\rm tape}=0.35f_{\rm peak}$, with $f_{\rm peak}$ the peak frequency of the $(2,2)$ mode~\cite{bohe_seobnrv4_2017}, $\gamma=1/50$ and $N_{\rm GW}=1$, following~\cite{mehta_fti_2023}. The same tapering frequency is used in the parametrized tests of~\cite{abbott_tests_2021}.

We use the modified waveform without linearization, so that its deviation from GR, $\delta h=h[\exp(i\beta\phi_b)-1]$, reduces to $i\beta h\phi_b$ only at linear order.

\subsection{Template and response function}
\label{sec:residual}

For noise with one-sided power spectral density $\Sn(f)$, we define
\begin{equation}
\langle a,b\rangle=4\int\frac{a^*(f)b(f)}{\Sn(f)}\,df,\qquad
\ip{a}{b}=\operatorname{Re}\langle a,b\rangle,\qquad
\norm{h}=\ip{h}{h}^{1/2}.
\label{eq:inner}
\end{equation}
Maximizing over a relative phase gives the magnitude of the complex overlap, so the match is
\begin{equation}
\mathcal{M}(h,d)=\max_{t_0}\frac{|\langle h_{t_0},d\rangle|}{\norm{h}\norm{d}},\qquad
h_{t_0}(f)=h(f)e^{2\pi ift_0},
\label{eq:match}
\end{equation}
where $d$ denotes the noisy data. The mismatch is $\bar{\mathcal{M}}=1-\mathcal{M}$.

For every realization, including GR samples, we keep the intrinsic parameters of $h$ fixed at their injected values and determine $t_0$ by maximizing the match in eq.~\eqref{eq:match}. At this time shift, the complex least-squares coefficient and the matched template are
\begin{equation}
a=\frac{\langle h_{t_0},d\rangle}{\langle h_{t_0},h_{t_0}\rangle},\qquad
h_{\rm m}=a h_{t_0},\qquad r=d-h_{\rm m}.
\label{eq:amplitude}
\end{equation}
The coefficient $a$ determines the amplitude and phase and makes $\langle h_{\rm m},r\rangle=0$, so the fit is restricted to the three extrinsic quantities and requires the GR waveform at the true intrinsic parameters.

Let us briefly recall how the response function is defined in~\cite{hemmatyar_paper1_2026}. For a scalar observable $O$ that depends on the waveform, such as the mismatch, the first-order change of $O$ under a phase deformation $\dpsi(f)$ is written as an integral over frequency,
\begin{equation}
\delta O=\operatorname{Re}\int R_O(f)\,\dpsi(f)\,df,
\label{eq:response_def}
\end{equation}
and the response function $R_O$ is the kernel of this integral. It is read off once the first-order variation of $O$ has been brought into the form of eq.~\eqref{eq:response_def}, and an analogous kernel describes amplitude deformations. In this work, the observable is the mismatch, and the derivation of its response function is given in appendix~\ref{app:response}. The variation of the mismatch under a small waveform change $\Delta h$ is
\begin{equation}
\delta\bar{\mathcal{M}}=-\frac{\ip{\Delta h}{d}}{\norm{h}\norm{d}}
+\mathcal{M}\frac{\ip{h}{\Delta h}}{\norm{h}^2}.
\label{eq:dmismatch}
\end{equation}
A phase deformation corresponds to $\Delta h=ih\dpsi$. Inserting it into eq.~\eqref{eq:dmismatch} and writing the inner products as frequency integrals with eq.~\eqref{eq:inner}, the variation of the mismatch takes the form of eq.~\eqref{eq:response_def}, $\delta\bar{\mathcal{M}}=\operatorname{Re}\int R(f)\dpsi(f)\,df$, with the mismatch response function
\begin{equation}
R(f)=\frac{4ih(f)}{\norm{h}\Sn(f)}
\left[-\frac{d^*(f)}{\norm{d}}+\mathcal{M}\frac{h^*(f)}{\norm{h}}\right].
\label{eq:response}
\end{equation}
The network uses both components of this complex response function. When $h$ is replaced by $h_{\rm m}$, eq.~\eqref{eq:amplitude} gives $\mathcal{M}=\norm{h_{\rm m}}/\norm{d}$ and therefore
\begin{equation}
R(f)=-\frac{4i h_{\rm m}(f)r^*(f)}{\norm{h_{\rm m}}\norm{d}\Sn(f)}.
\label{eq:response_used}
\end{equation}
This identity uses the least-squares amplitude in eq.~\eqref{eq:amplitude}, since the normalized match does not fix the template magnitude, and expresses the response function through the waveform residual. The product with the conjugate residual removes the rapid common phase of the signal and template, so that coherent deviations become visible directly in frequency.

We reduce $R$ to 512 logarithmically spaced frequency cells between $f_{\rm low}$ and $400$~Hz, where each cell contains the mean of $R$ over the frequencies of the waveform grid that fall inside it. The real and imaginary components form two input channels. Each is divided by its own standard deviation over the cells below $\min(f_{\rm peak},400\,{\rm Hz})$ and set to zero above. Consequently, the overall factors in eq.~\eqref{eq:response_used} drop out and the separate channel scales are discarded.

\subsection{Sampling protocol}
\label{sec:protocol}

The same value of the coupling $\beta$ can produce very different waveform residuals for different exponents and sources. We therefore draw samples using a coordinate defined by the residual norm. Let $\pperp$ be the orthogonal projector, with the real inner product in eq.~\eqref{eq:inner}, onto the complement of the three directions $h$, $ih$ and $ifh$. These are the local amplitude, phase and time directions. We define
\begin{equation}
s=\norm{\pperp\delta h(\beta)},\qquad
\delta h(\beta)=h\left[e^{i\beta\phi_b(f)}-1\right].
\label{eq:s}
\end{equation}
The coordinate is symmetric under changing the sign of $\beta$, and we invert it numerically on the monotonic branch of a scan in $|\beta|$.

For each source and each of its 100 noise realizations, we construct seven samples, one GR sample and one sample for each exponent, all containing the same noise. In this way, the differences between the seven samples are due to the deviations alone, and each class contains the same number of samples. For each exponent, we draw $s$ log-uniformly between $0.5$ and an upper limit $s_{\rm cap}$, draw the sign of $\beta$ with equal probability, and obtain $|\beta|$ by inverting eq.~\eqref{eq:s}. The lower limit corresponds to a residual well below the noise level, and the log-uniform distribution gives equal weight to each decade of strength. The upper limit $s_{\rm cap}$ follows from three requirements, and for each source and exponent we take the most restrictive of them. First, we keep the deviation small compared with the signal by requiring $s\le\mathrm{SNR}/3$, so that the residual carries at most one ninth of the signal power and the linear definition of $s$ describes the residual that is actually present. This factor is a choice, which affects the aggregate accuracies and determines which sources contribute at large $s$. Second, the few loudest sources, with SNRs up to $137$, should not alone populate the strongest residuals, and we therefore also require $s\le20$. Third, the value of $s$ must determine $|\beta|$ uniquely. As $|\beta|$ grows, the phase correction wraps around and $s$ stops increasing, so that the same value of $s$ can be reached with more than one coupling. We therefore remain below the largest value $s_{\rm max}$ reached while $s$ still increases, with a margin of $2\%$, since $s$ hardly changes with $|\beta|$ near this value. Altogether, $s_{\rm cap}=\min(20,\mathrm{SNR}/3,0.98\,s_{\rm max})$, where the last bound is rarely active.

A source is kept only if all six exponents produce a visible deviation in the detector band. This requires that the taper frequency lies above the lower cutoff, since otherwise the correction vanishes before it enters the band, and that every exponent reaches $s=1$. The Hanford row also requires an SNR of at least $8$. These conditions keep all 262 aLIGO sources and 132 Hanford sources.

The coordinate $s$ is the residual norm that the match would leave if the time shift were linear, $e^{2\pi ift_0}\simeq1+2\pi ift_0$. The actual match maximizes over the full time shift, and we denote by $s_{\rm exact}$ the norm of the residual $r$ in eq.~\eqref{eq:amplitude} for the modified signal without noise. We draw samples in $s$, which can be computed and inverted for $\beta$ before any noise is added, and present all strength-dependent results in terms of $s_{\rm exact}$, the strength actually present in the data.

Each source thus contributes 700 samples, seven for each of its 100 noise realizations, so that the aLIGO and Hanford datasets contain $183{,}400$ and $92{,}400$ samples. We split the sources into training, validation and test sets with fractions of about $70\%$, $15\%$ and $15\%$. The test sets therefore contain 40 aLIGO and 19 Hanford sources, and no template or noise realization of a test source enters training. Fixed seeds allow the reference to be evaluated on the same realizations as the network.

\section{Neural network and evaluation}
\label{sec:classifiers}

We now introduce the network with its two output heads. We then define the residual-bank reference and the quantities used to assess the classification, the detection of deviations and the regression of the coupling.

\subsection{Network architecture and training}
\label{sec:network}

\begin{figure}[t]
\centering
\includegraphics[width=\textwidth]{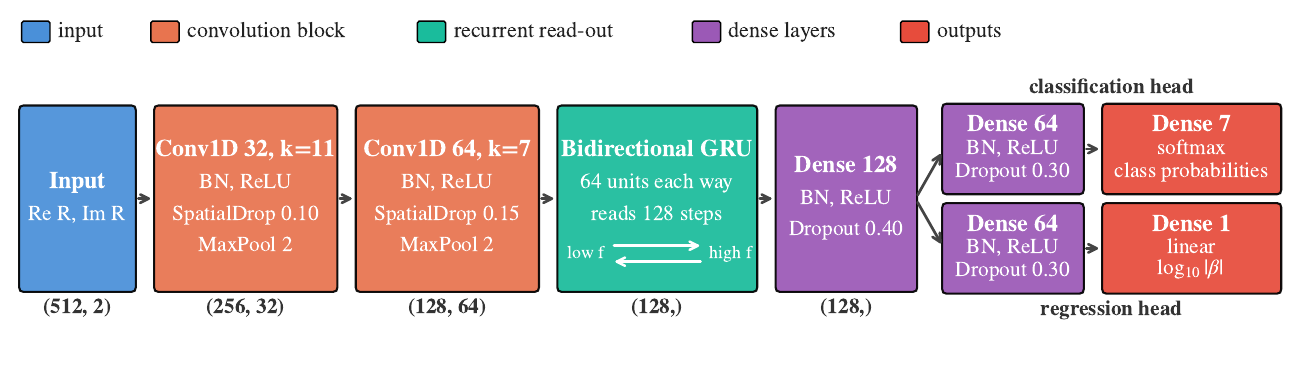}
\caption{Architecture of the network. The real and imaginary parts of the response function on 512 frequency cells pass through two convolution blocks and a bidirectional GRU, which reads the resulting 128 feature vectors along frequency in both directions. Its final states feed a shared dense layer and two heads, which output the class probabilities and the standardized $\log_{10}|\beta|$. The shapes below the blocks give the number of frequency cells and channels, or the number of units. BN denotes batch normalization.}
\label{fig:architecture}
\end{figure}

The network contains two convolution blocks. The first has 32 filters of width 11 and the second 64 filters of width 7. Each convolution is followed by batch normalization~\cite{ioffe_batchnorm_2015}, a ReLU activation, spatial dropout~\cite{tompson_spatialdropout_2015,srivastava_dropout_2014} and max-pooling by two. The dropout rates are $0.10$ and $0.15$. A bidirectional gated recurrent unit~\cite{cho_gru_2014,schuster_bidirectional_1997}, with 64 units in each direction, reads the remaining 128 frequency steps. The final states feed a shared dense layer with 128 units and two heads with 64 hidden units each. These dense hidden layers also use batch normalization and ReLU activations, with dropout $0.4$ in the shared layer and $0.3$ in each head. The classification output is a softmax over seven classes. The regression output is a single linear unit for the standardized $\log_{10}|\beta|$, with standardization determined from the training targets. Figure~\ref{fig:architecture} shows the architecture.

The model has $100{,}008$ parameters. An $L_2$ penalty of $5\times10^{-4}$ is applied to the convolutional and hidden dense kernels. The recurrent read-out retains the ordering of features along the band, whereas global average pooling summarizes their average presence. Convolutional and recurrent layers have also been combined for sequence classification in other settings~\cite{sainath_cldnn_2015,shi_crnn_2017,ordonez_deepconvlstm_2016}. We investigate the choice empirically in section~\ref{sec:variants}.

The loss is categorical cross-entropy with label smoothing $\epsilon=0.05$~\cite{szegedy_rethinking_2016}, plus $\lambda=0.1$ times the squared error of the regression head. The regression loss is masked on GR samples. We train with Adam~\cite{kingma_adam_2015}, an initial learning rate of $3\times10^{-4}$ and batches of 256, using Keras and TensorFlow~\cite{chollet_keras_2015,abadi_tensorflow_2016}. The learning rate is halved after five epochs without improvement in validation loss. Early stopping uses a patience of ten epochs and restores the best weights, with a cap of 80 epochs. Each main construction is trained with three seeds.

\subsection{Reference classifier}
\label{sec:reference}

The accuracy of the network alone does not tell whether a low value is due to the network or to the problem itself, that is, whether the phase families can be distinguished in the data at all. In order to separate the two, we construct a reference classifier that uses the physics of the problem directly and is evaluated on the same samples as the network.

The starting point is the residual $r$ of eq.~\eqref{eq:amplitude}, which remains after the match. For a GR sample it contains only noise, whereas for a sample of family $b$ it contains, in addition, the part of the deviation that the match does not absorb. Since the parameters of each source are known, this surviving part can be computed in advance without noise. For each source, exponent and sign, we denote by $m_b(s)$ the noise-free residual left by the match at strength $s$, compute it on a logarithmic grid of 48 strengths between $0.5$ and $s_{\rm cap}$, and interpolate between the grid points. The collection of these residuals plays the role of a template bank for residuals, and we therefore refer to this classifier as the residual bank.

In order to decide which family best explains a given residual, we compare it with the bank through
\begin{equation}
\ell_b=\ip{r}{m_b}-\frac{1}{2}\norm{m_b}^2=\frac{1}{2}\norm{r}^2-\frac{1}{2}\norm{r-m_b}^2.
\label{eq:bank_score}
\end{equation}
The interpretation of this quantity is rather simple: $\norm{r-m_b}$ is the distance between the residual and the bank residual, while $\norm{r}$ is its distance from zero, which is the residual expected in GR. Hence $\ell_b$ is positive when the residual lies closer to $m_b$ than to GR. For Gaussian noise, $\ell_b$ is precisely the logarithm of the likelihood ratio between the hypothesis $r=m_b+n$ and GR, for which $r=n$.

The strength and the sign of a deviation are not known for a given sample. For each exponent, we therefore average $\exp(\ell_b)$ over both signs and over $s$ with the log-uniform distribution used to generate the samples, which gives one score per exponent. GR corresponds to $m_b=0$ and hence to the score one. The reference assigns each sample to the class with the largest score.

If the likelihood in eq.~\eqref{eq:bank_score} were exact, this rule would be the best possible classifier for our samples, since it would compare the exact probabilities of the seven classes. It is approximate, because the time match depends on the noise, so that the residual of a noisy sample is not exactly the noise-free residual plus Gaussian noise. The residual bank is therefore a physically motivated benchmark rather than a strict upper bound. Notice also that it uses more information than the network, since it is built from the parameters of each source and keeps the absolute size of the residual, which the channel normalization removes from the network input. Consequently, if the network reaches an accuracy close to that of the residual bank, it extracts most of the information available to this benchmark, while a low accuracy of both indicates that the phase families themselves are difficult to distinguish.

We use two further comparisons. The first is a simplified residual bank, which we call the linear-ray reference. For small couplings, the residual of family $b$ grows along a fixed direction $\hat u_b$, given in eq.~\eqref{eq:geometry}, so that $m_b(s)\simeq\pm s\,\hat u_b$. As $s$ increases, these approximate residuals move from zero along a straight line, a ray, and the linear-ray reference uses them in place of the exact residuals. Comparing it with the residual bank shows how much the classification relies on the nonlinear shape of the residuals. The second applies principal-component analysis (PCA) followed by linear discriminant analysis (LDA)~\cite{pedregosa_sklearn_2011} to the network inputs, as a simple linear classifier on the same data as the network, and is described in section~\ref{sec:unsupervised}.

\subsection{Metrics and uncertainty}
\label{sec:metrics}

We report seven-class accuracy over the full test set, accuracy on loud beyond-GR samples with $s_{\rm exact}>5$, and GR recall at the largest-score decision. Since the strengths are drawn log-uniformly, a large fraction of the beyond-GR samples carry weak deviations that are hidden by the noise and are rarely identified by any classifier. The overall accuracy is therefore low and depends on how many weak samples the strength distribution contains, whereas the loud subset shows how well the families are identified when the deviation is clearly present.

We also test the classifiers as tools for searching for deviations. Here the main risk is to flag a GR signal as a deviation. A sample is flagged when its probability of being GR is low, and we set the threshold such that $1\%$ of the GR samples in the validation set are flagged. On the test set, we then measure three fractions: the GR samples that are flagged, the beyond-GR samples that are flagged, which we call the efficiency, and the beyond-GR samples that are flagged with the correct exponent, which we call the identification. Finally, we check whether the class probabilities of the network can be trusted, that is, whether a probability of $0.7$ means a correct decision in about $70\%$ of the cases, using the expected calibration error~\cite{guo_calibration_2017}.

The test samples are not independent, since all samples of a source share its parameters and the seven samples of a realization share their noise. We therefore estimate uncertainties with a bootstrap over sources. We draw the test sources with replacement $10{,}000$ times, recompute each quantity from the samples of the drawn sources, and quote the spread of the results. For recorded noise, we also redraw the 4096-s segments of data from which the noise stretches are taken. When two classifiers are compared, both are evaluated on the same draws, and for the network the decisions of the three seeds are first averaged for each sample.

To describe the confusion at a coarser resolution, we also merge the seven-way decisions into GR, $b<0$ and $b>0$, and count the fraction of loud samples assigned to their true exponent or an adjacent exponent in the ordered class list. These are post-hoc summaries of the same predictions.

\section{Geometry of the deviations}
\label{sec:geometry}

In order to understand which phase families are difficult to distinguish, we compare the residuals that they leave after the match. For small couplings, the deviation of family $b$ is $\delta h\simeq i\beta h\phi_b$, and the part that survives the match is its projection with $\pperp$ from section~\ref{sec:protocol}. Every family therefore defines a direction in the space of residuals,
\begin{equation}
\hat u_b=\frac{\pperp(ih\phi_b)}{\norm{\pperp(ih\phi_b)}},
\label{eq:geometry}
\end{equation}
and a deviation of strength $s$ in family $b$ leaves the residual $\pm s\,\hat u_b$, with the sign of the coupling. Two families are difficult to distinguish when their directions are close. Since both signs of the coupling are admitted, we measure this by the angle
\begin{equation}
\theta_{bc}=\arccos\left|\ip{\hat u_b}{\hat u_c}\right|,
\label{eq:angle}
\end{equation}
which is small when the residuals of the two families are nearly parallel or nearly antiparallel. For two residuals of equal strength $s$ and a small angle, the distance between them is about $s\,\theta_{bc}$, with $\theta_{bc}$ in radians, so that a small angle requires a strong deviation to separate the two families.

\begin{figure}[t]
\centering
\includegraphics[width=\textwidth]{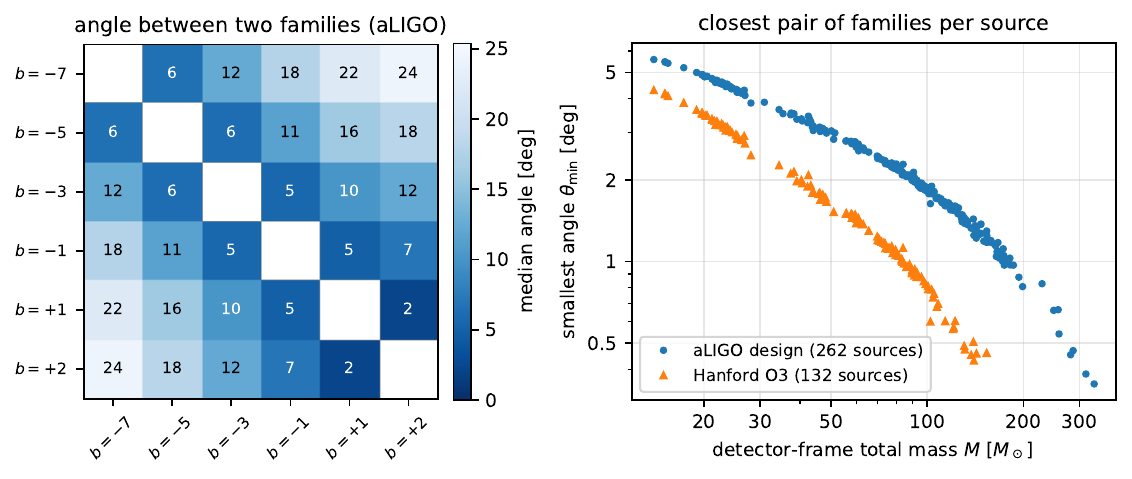}
\caption{Similarity of the residual directions of the phase families after the match. The angle $\theta_{bc}$ between two families is defined in eq.~\eqref{eq:angle}, and a small angle means that the two families leave nearly the same residual. Left: median angle over the 262 aLIGO sources for every pair of exponents. Right: the smallest of these angles for each source, against the detector-frame total mass, for aLIGO and Hanford O3. All six families leave nearly parallel residuals, and neighboring exponents differ by only a few degrees, which explains why the classifiers mostly confuse neighboring families and why heavy sources are the most difficult.}
\label{fig:geometry}
\end{figure}

Figure~\ref{fig:geometry} shows the result. The left panel gives the median angle between every pair of families on aLIGO. All six families lie within a narrow cone of about 25 degrees, and the angles grow along the ladder of exponents: neighboring families differ by only a few degrees, while distant families differ more. The closest pair is $b=+1$ and $b=+2$, and it is the closest pair for every source. This pattern mirrors the confusion of the classifiers in figure~\ref{fig:confusion}, where most errors fall on neighboring exponents. The right panel shows the smallest angle for each source. It decreases with the mass of the source, since a heavier source has a shorter inspiral in the detector band, over which the phase families look more alike. It is also smaller on Hanford, whose lower cutoff of $20$~Hz shortens the band further.

These directions describe the residuals at linear order. For strong deviations, the exact residuals $m_b(s)$ used by the residual bank bend away from their linear directions, and this nonlinear structure helps to distinguish the families. Indeed, on loud samples the linear-ray reference, which uses the directions of eq.~\eqref{eq:geometry}, identifies the exponent far less often than the residual bank, as shown in table~\ref{tab:headline}.

\FloatBarrier
\section{Results on Gaussian noise}
\label{sec:results}

We now present the results on Gaussian noise. We first compare the accuracy of the network with that of the residual bank and show how it depends on the strength of the deviation and on the mass of the source. We then examine which families are confused with each other. Next, we use the network to flag deviations with a fixed fraction of false positives, discuss the regression of the coupling, and finally examine the network inputs with simple linear methods.

\subsection{Classification and comparison with the reference}
\label{sec:headline}

\begin{table}[t]
\centering\footnotesize\setlength{\tabcolsep}{4pt}
\begin{tabular}{llcccc}
\toprule
row & classifier & 7-class & boot. s.d. & loud & GR recall \\
\midrule
aLIGO & network & 0.254 $\pm$ 0.001 & 0.007 & 0.427 $\pm$ 0.030 & 0.819 \\
aLIGO & residual bank & 0.267 & 0.008 & 0.469 $\pm$ 0.034 & 0.806 \\
aLIGO & linear rays & 0.226 & 0.007 & 0.241 $\pm$ 0.022 & 0.819 \\
Hanford O3 & network & 0.218 $\pm$ 0.001 & 0.006 & 0.329 $\pm$ 0.014 & 0.821 \\
Hanford O3 & residual bank & 0.226 & 0.007 & 0.405 $\pm$ 0.021 & 0.758 \\
Hanford O3 & linear rays & 0.203 & 0.007 & 0.197 $\pm$ 0.023 & 0.788 \\
\bottomrule
\end{tabular}
\caption{Accuracy of the network and of the two references on the same test samples. Network values are means over three seeds. The value after $\pm$ is the spread over seeds for the seven-class accuracy and the source-bootstrap standard deviation for the loud accuracy, and the column ``boot.\ s.d.'' gives the source-bootstrap standard deviation of the seven-class accuracy. The loud subsets contain 1990 samples from 21 aLIGO sources and 259 samples from four Hanford sources.}
\label{tab:headline}
\end{table}

Table~\ref{tab:headline} gives the accuracies of the network and of the two references on the same test samples. Over all samples, the network lies about one percentage point below the residual bank on both rows, and the source-bootstrap intervals of these differences exclude zero. On loud samples, the network identifies the exponent in $0.427$ of the cases on aLIGO and in $0.329$ on Hanford, compared with $0.469$ and $0.405$ for the residual bank. The Hanford loud subset contains only four sources, so that its values are less certain.

\begin{figure}[t]
\centering
\includegraphics[width=0.9\textwidth]{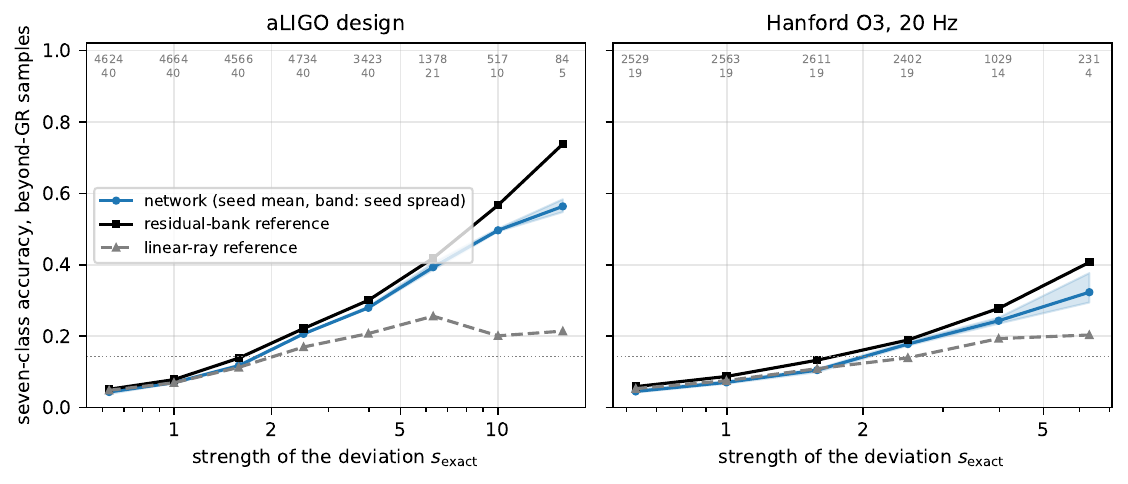}
\caption{Accuracy on beyond-GR test samples against the strength $s_{\rm exact}$ of the deviation. The network curve is the mean over three seeds, and the band spans the seeds. The annotations give the number of samples and sources in each bin, and bins with fewer than 50 samples are omitted. The dotted line marks random guessing among seven classes.}
\label{fig:accuracy_vs_s}
\end{figure}

Figure~\ref{fig:accuracy_vs_s} shows how the accuracy grows with the strength of the deviation. For weak deviations, all classifiers assign most beyond-GR samples to GR, since the deviation is hidden in the noise. As the deviation becomes stronger, the accuracy rises steadily. The network follows the residual bank for weak and moderate deviations, and falls behind it for the strongest ones, which come from only a few sources. The linear-ray reference stops improving at large strength, where the exact residuals bend away from their linear directions (section~\ref{sec:geometry}).

The accuracy also depends on the mass of the source. On loud aLIGO samples, the residual bank identifies the exponent in $61\%$ of the cases for sources below $30\,M_\odot$ and in $37\%$ above $60\,M_\odot$, and the network in $53\%$ and $34\%$. This is expected from section~\ref{sec:geometry}, since a heavier source has a shorter inspiral in the detector band, where the families lie closer to each other.

The two classifiers also fail on the same samples. On both rows, about $90\%$ of the wrong decisions of the network occur on samples that the bank also misclassifies, and in most of these cases the two choose the same wrong class. Finally, the main networks show no sign of overfitting, since the accuracy on the training set exceeds that on the test set by at most about one percentage point, and the best validation loss is reached before the maximum of 80 epochs.

\subsection{Confusion and coarse groups}
\label{sec:confusion}

\begin{figure}[t]
\centering
\includegraphics[width=0.66\textwidth]{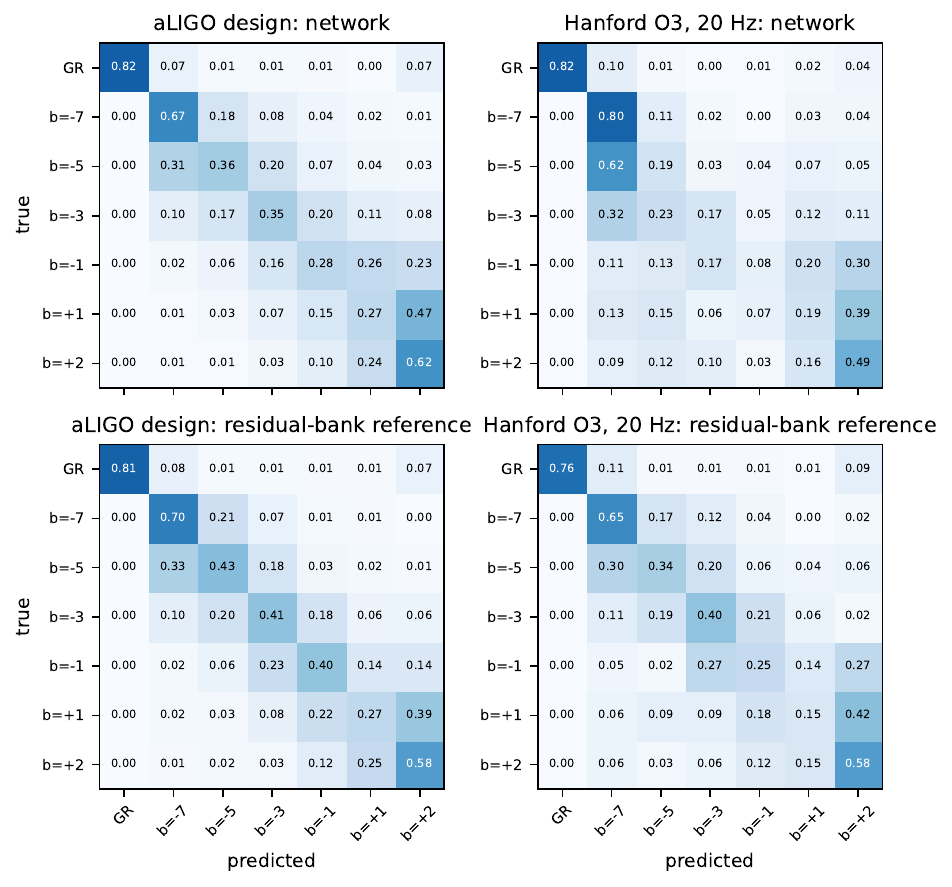}
\caption{Confusion matrices of the network, with the decisions of the three seeds pooled, and of the residual bank. Each row shows how the samples of one true class are distributed over the predicted classes. The beyond-GR rows contain only loud samples, and the GR row contains all GR test samples.}
\label{fig:confusion}
\end{figure}

Figure~\ref{fig:confusion} shows which families are confused with each other. The errors follow the ladder of exponents found in section~\ref{sec:geometry}. Most wrong decisions fall on a neighboring exponent, and the confusion is largest among the four intermediate exponents, which have neighbors on both sides. The two outer families, $b=-7$ and $b=+2$, are identified most often, in more than $60\%$ of the loud aLIGO samples. The confusion is, however, not restricted to neighbors, since about a third of the wrong loud decisions on aLIGO, and about half on Hanford, fall further away.

\begin{table}[t]
\centering\footnotesize\setlength{\tabcolsep}{4pt}
\begin{tabular}{llccccc}
\toprule
row & classifier & three-way & recall $b<0$ & recall $b>0$ & within one & exact class \\
\midrule
aLIGO & network & 0.810 & 0.814 & 0.799 & 0.813 & 0.427 \\
aLIGO & residual bank & 0.848 & 0.896 & 0.742 & 0.855 & 0.469 \\
aLIGO & linear rays & 0.598 & 0.455 & 0.916 & 0.550 & 0.241 \\
Hanford O3 & network & 0.740 & 0.781 & 0.621 & 0.673 & 0.329 \\
Hanford O3 & residual bank & 0.803 & 0.855 & 0.652 & 0.776 & 0.405 \\
Hanford O3 & linear rays & 0.598 & 0.503 & 0.879 & 0.475 & 0.197 \\
\bottomrule
\end{tabular}
\caption{Coarser summaries of the same decisions on loud beyond-GR samples, without retraining. The three-way accuracy merges the classes into GR, $b<0$ and $b>0$, and the next two columns give the recall of the two merged groups. The within-one column counts decisions at the true exponent or a neighboring one. Network values are means over three seeds.}
\label{tab:groups}
\end{table}

Since neighboring families are hard to separate, we also ask coarser questions of the same decisions, as summarized in table~\ref{tab:groups}. If the classes are merged into GR, negative exponents and positive exponents, the network assigns about $80\%$ of the loud aLIGO samples to the correct group, and the same fraction lies within one neighbor of the true exponent. The values on Hanford are lower. A coarse description of a deviation is therefore recovered much more reliably than its exact exponent. We chose these groupings after inspecting the confusion matrices.

\begin{figure}[t]
\centering
\includegraphics[width=0.72\textwidth]{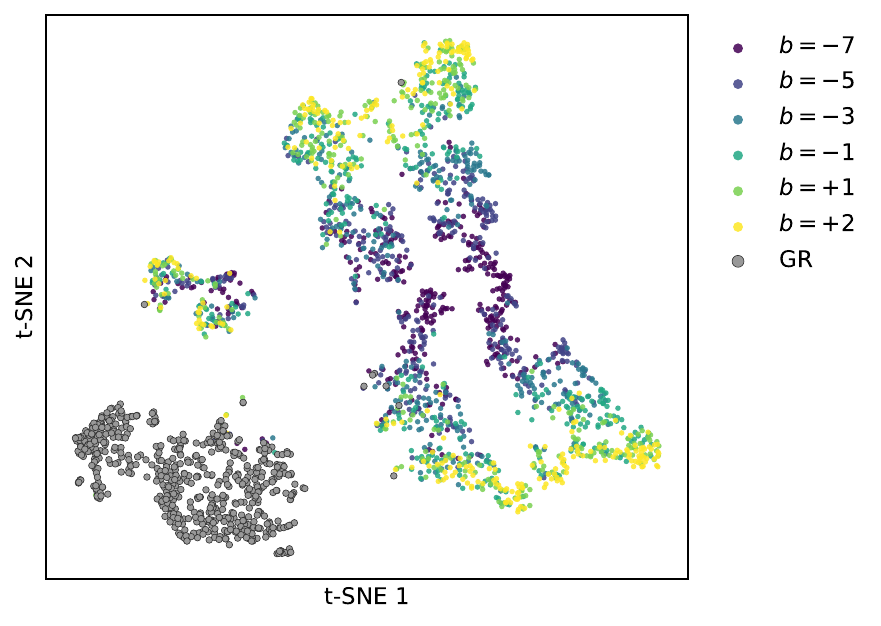}
\caption{t-SNE map of the shared-layer activations for 600 GR samples and all loud deviations of the aLIGO test set, for one seed. Nearby points are represented similarly by the network. The GR samples, in gray, form a separate cluster, while the six families are ordered along the ladder of exponents and overlap with their neighbors.}
\label{fig:tsne}
\end{figure}

We can also look at how the network represents the samples internally. Figure~\ref{fig:tsne} shows a t-SNE map~\cite{vandermaaten_tsne_2008} of the shared layer for GR samples and loud deviations on aLIGO, which is a two-dimensional picture in which samples with similar internal representations lie close together. Two features stand out. First, the GR samples form a cluster of their own, separate from all loud deviations, which is why nearly every loud deviation is detected. Second, the six families are not separated, but they are ordered. The map runs from $b=-7$ through the intermediate exponents to $b=+2$, and neighboring families overlap. This is the ladder of section~\ref{sec:geometry}, seen from inside the network, and it explains why most errors fall on a neighboring exponent. The few separate islands in the map contain the samples of individual sources.

\subsection{Detection and calibration}
\label{sec:operating}

\begin{table}[t]
\centering\footnotesize\setlength{\tabcolsep}{3pt}
\begin{tabular}{llccccc}
\toprule
row & classifier & GR false-pos. [\%] & eff. & ident. & loud eff. & loud ident. \\
\midrule
aLIGO & network & 0.83 $[0.61,1.04]$ & 0.304 & 0.093 & 0.993 & 0.425 $\pm$ 0.030 \\
aLIGO & residual bank & 0.97 $[0.70,1.27]$ & 0.336 & 0.109 & 0.997 & 0.468 $\pm$ 0.034 \\
Hanford O3 & network & 1.35 $[0.88,1.88]$ & 0.240 & 0.056 & 0.994 & 0.328 $\pm$ 0.014 \\
Hanford O3 & residual bank & 1.21 $[0.58,2.00]$ & 0.285 & 0.070 & 1.000 & 0.405 $\pm$ 0.021 \\
\bottomrule
\end{tabular}
\caption{Detection of deviations when the threshold of each model flags $1\%$ of its GR validation samples. The false-positive column gives the fraction of GR test samples that are flagged, with its source-bootstrap $95\%$ interval. Efficiency and identification are the fractions of beyond-GR samples that are flagged, and that are flagged with the correct exponent. Loud identification carries its source-bootstrap standard deviation. Network values are means over three seeds.}
\label{tab:operating}
\end{table}

Table~\ref{tab:operating} separates two questions, namely whether a deviation is detected and whether its exponent is identified. With a threshold that flags $1\%$ of the GR validation samples, the fraction of flagged GR test samples stays close to $1\%$ on both rows. Nearly every loud deviation is flagged, by both the network and the bank. However, fewer than half of the loud deviations are flagged with the correct exponent. A loud deviation is thus almost always detected, while its family is identified much less often. Over all strengths the efficiency is much lower, since most weak deviations cannot be told apart from GR.

The class probabilities of the network are also well calibrated for the distribution of our samples. Their expected calibration error is about one percent or less for every seed, comparable with that of the bank.

\subsection{The auxiliary regression head}
\label{sec:regression}

The second head of the network is trained to estimate the magnitude of the coupling, $\log_{10}|\beta|$, of beyond-GR samples. This task is harder than it may appear, because the couplings of different families lie in very different ranges. The correction for $b=-7$ grows as $f^{-7/3}$ towards low frequencies, so that a very small coupling already produces a strong deviation, whereas $b=+2$ needs a much larger coupling for the same strength. In our samples, the couplings therefore span more than ten orders of magnitude, and each family occupies its own part of this range.

\begin{figure}[t]
\centering
\includegraphics[width=0.52\textwidth]{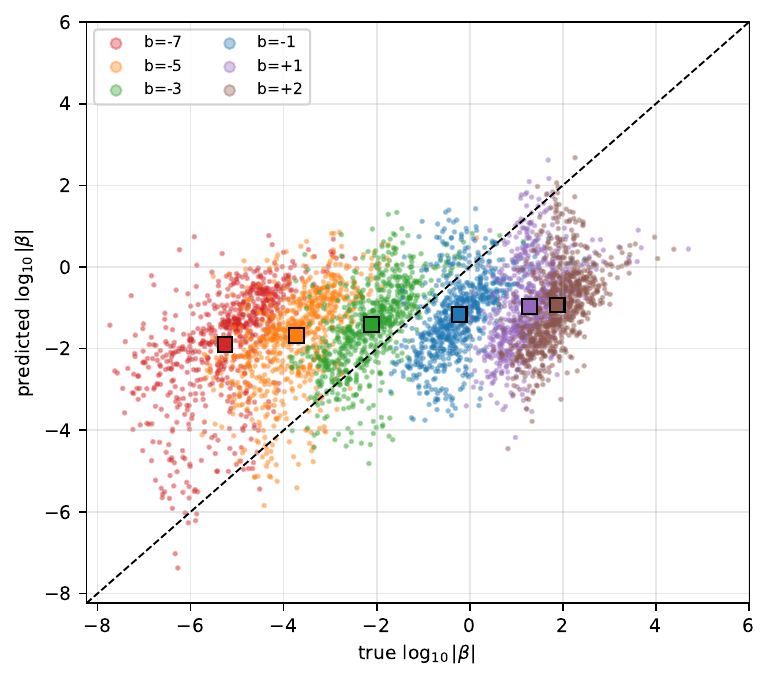}
\caption{Predicted against true $\log_{10}|\beta|$ for beyond-GR aLIGO test samples of one seed. Each color is one family, and the squares mark the average prediction of each family. A perfect estimate would lie on the diagonal, whereas the squares lie on a nearly flat line, which shows that the predictions are pulled towards the center of the range.}
\label{fig:regression}
\end{figure}

Figure~\ref{fig:regression} compares the predicted and the true $\log_{10}|\beta|$. A perfect estimate would place all points on the diagonal. Instead, the average predictions of the six families, shown as squares, lie on a nearly flat line, so that all families are pulled towards the center of the range. The head therefore overestimates the small couplings of $b=-7$ and underestimates the large couplings of $b=+2$, on average by about three orders of magnitude. Within a family, the predictions form a broad cloud that follows the true coupling only weakly, and for every family they are worse than simply using the average coupling of that family. Over all samples, the head is only slightly better than a single constant, and only slightly better than a guess that uses nothing but the class probabilities of the network, weighting the average coupling of each family by its probability. The small gain of the head therefore comes from recognizing the family, rather than from measuring the coupling.

Two features of our setup explain this behavior. First, within a family the coupling of a sample is set by the strength of its deviation, and the channel normalization of the input removes the overall size of the response function, which is the most direct measure of this strength. Second, when the family is uncertain, a single output trained with a squared error is pulled towards an average over the possible families, which lies near the center of the range. We therefore regard the second head as an auxiliary output rather than an estimator of the coupling. An estimate that keeps the size of the response function and is conditioned on the predicted family is left for future work.

\subsection{PCA and a linear baseline}
\label{sec:unsupervised}

Finally, we ask whether the classes can be separated by simple linear methods applied directly to the network inputs. We use two standard tools~\cite{pedregosa_sklearn_2011}. Principal component analysis (PCA) finds the directions along which the inputs vary most, without using the class labels. Linear discriminant analysis (LDA), in contrast, uses the labels to find the directions that best separate the classes, and can then be used as a linear classifier.

\begin{figure}[t]
\centering
\includegraphics[width=\textwidth]{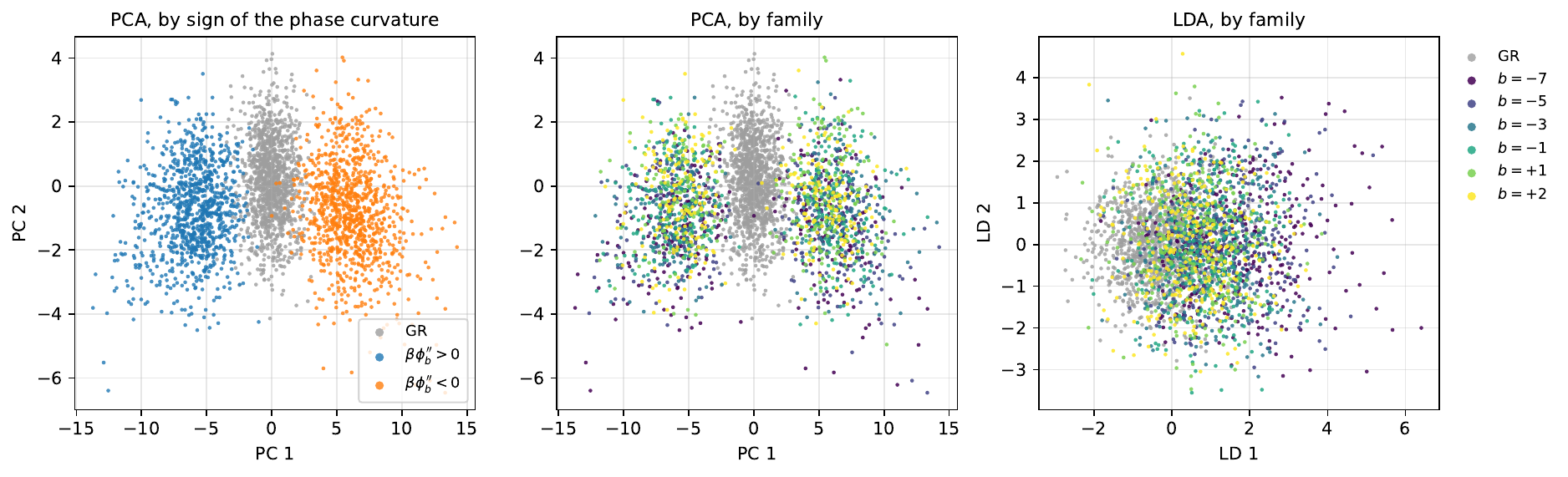}
\caption{Linear projections of the aLIGO response functions for GR samples, in gray, and loud deviations. Left and middle: the two leading principal components, colored by the sign of the phase curvature $\beta\phi_b''$ and by family. Right: the two leading axes of LDA trained on the 50 leading principal components. The leading principal components separate the deviations from GR and split them by the sign of the phase curvature, but not by family, and even LDA separates the families only weakly.}
\label{fig:unsupervised}
\end{figure}

Each input consists of 1024 numbers, the two channels on 512 frequency cells, and most of its variation is noise. We therefore first reduce the inputs with PCA to their 50 leading components, computed from $30{,}000$ training samples, and then train LDA on these components. Figure~\ref{fig:unsupervised} shows the result for aLIGO. In the two leading principal components, loud deviations form two groups on either side of GR. The two groups correspond to the two signs of $\beta\phi_b''$, the curvature of the phase correction, which determines whether the deviation speeds up or slows down the frequency sweep of the signal, and each group contains all families. This curvature changes sign between negative and positive exponents, so that the two groups mix the sign of the coupling with the sign of $b$. Even LDA, which uses the family labels, separates the families only weakly and reaches clearly lower accuracies than the network, for instance $0.20$ instead of $0.43$ on loud aLIGO samples. The network therefore uses structure in the response functions that this linear classifier does not capture.

\FloatBarrier
\section{Injections into recorded Hanford O3 strain}
\label{sec:realnoise}

In order to test whether the network works in real detector noise, we inject the Hanford signals into recorded O3 strain. Recorded noise exists only at the sensitivity of the run in which it was taken, and the aLIGO design sensitivity has not been reached in the observing runs so far. We therefore use the Hanford row, whose Gaussian dataset has the representative O3 spectrum and is thus the Gaussian counterpart of the recorded data. We train networks on Gaussian noise and on recorded noise, and test each of them on both.

We use stretches of 64~s from 17 segments of 4096~s of Hanford O3 data provided by GWOSC~\cite{gwosc,abbott_o3opendata_2023}, avoiding known events, times with data-quality or hardware-injection flags, and stretches with excess noise power. The noise spectrum of each stretch is estimated from the surrounding 1024~s with the median Welch method~\cite{welch_1967,allen_findchirp_2012}.

We train and validate the networks on data from the first part of the run, O3a, and test them on data from the second part, O3b, so that the test noise is never seen during training. Apart from the noise and the spectrum used in the response function, the recorded-noise dataset is identical to the Gaussian Hanford dataset, so that the two can be compared sample by sample.

\begin{table}[htb]
\centering\scriptsize\setlength{\tabcolsep}{3pt}
\begin{tabular}{llcccccc}
\toprule
trained on & tested on & 7-class & loud & GR recall & GR false-pos. [\%] & eff. & loud ident. \\
\midrule
Gaussian & Gaussian & 0.217 $\pm$ 0.001 & 0.313 & 0.807 & 1.82 $[1.19,2.51]$ & 0.250 & 0.313 $\pm$ 0.015 \\
Gaussian & O3b strain & 0.209 $\pm$ 0.003 & 0.292 & 0.827 & 1.56 $[0.75,2.73]$ & 0.203 & 0.292 $\pm$ 0.075 \\
O3a strain & Gaussian & 0.217 $\pm$ 0.002 & 0.308 & 0.760 & 1.30 $[0.86,1.81]$ & 0.223 & 0.308 $\pm$ 0.011 \\
O3a strain & O3b strain & 0.210 $\pm$ 0.002 & 0.290 & 0.806 & 1.11 $[0.54,1.86]$ & 0.199 & 0.285 $\pm$ 0.077 \\
\bottomrule
\end{tabular}
\caption{Hanford networks trained on Gaussian or recorded noise and tested on both. The accuracies hardly depend on the kind of noise, and the fraction of wrongly flagged GR samples stays between about $1\%$ and $2\%$. Values are means over three seeds, with the spread over seeds beside the seven-class accuracy. The false-positive fraction is that of GR test samples flagged at a threshold that flags $1\%$ of the GR validation samples of the training noise, with its bootstrap $95\%$ interval, and loud identification carries its bootstrap standard deviation. The bootstrap resamples sources, and for recorded test noise also the data segments. The loud subsets contain 259 samples from four sources in Gaussian noise and 275 samples from five sources in recorded noise.}
\label{tab:realnoise}
\end{table}

The networks of this comparison are trained separately from those of section~\ref{sec:results}, which explains the small differences between the Gaussian values of tables~\ref{tab:headline} and~\ref{tab:realnoise}. Table~\ref{tab:realnoise} shows that the accuracies hardly change between the two kinds of noise. Whether trained on Gaussian or on recorded noise, the networks reach almost the same overall accuracy on recorded O3b data as on Gaussian noise, and their loud accuracies differ by at most about two percentage points. Since the strength of a sample is measured with the spectrum of its own noise, the loud samples are not exactly the same in the two kinds of noise.

We also check how often GR samples are wrongly flagged as deviations, with the threshold set to flag $1\%$ of the GR validation samples. On the test samples, this fraction lies between about $1\%$ and $2\%$ in all four cases. On recorded O3b noise, it is $1.6\%$ for the networks trained on Gaussian noise and $1.1\%$ for those trained on recorded noise. Training on recorded noise may therefore reduce the false alarms, but with the limited test data this difference is not yet significant.

\FloatBarrier
\section{Exploration of network variants}
\label{sec:variants}

In order to test whether the recurrent architecture is a good choice, we train thirteen variants on the aLIGO dataset, with two seeds each, and compare them with the recurrent network. Most variants modify a pooled network, which replaces the recurrent read-out by a third convolution block with 128 filters of width 5, followed by global average pooling. Global average pooling keeps only the average of each feature over the band, whereas the recurrent read-out keeps the order of the features in frequency. The variants are the following.
\begin{itemize}[leftmargin=1.6em,itemsep=2pt]
\item Recurrent networks: the network of section~\ref{sec:network} retrained with two seeds, a version with a third convolution block before the GRU, and a version trained for up to 160 epochs with a cosine learning-rate schedule and an early-stopping patience of 30 epochs.
\item Pooled networks: with twice as many filters, with the full feature map instead of its average, trained for up to 200 epochs with the same schedule, without the regression loss ($\lambda=0$), with half the dropout and no label smoothing, and with the frequency position as a third input channel.
\item Pooled networks with additional inputs through a small dense branch: the source parameters, namely chirp mass, symmetric mass ratio, spin, SNR and total mass, the scale of the real part of $R$ removed by the channel normalization, or both.
\item A residual network~\cite{he_resnet_2016} with about 2.2 million parameters, consisting of a first convolution and four stages of two residual blocks with 32, 64, 128 and 256 filters, followed by global average pooling.
\end{itemize}

\begin{table}[t]
\centering\footnotesize\setlength{\tabcolsep}{3pt}
\begin{tabular}{lcccccc}
\toprule
variant & seeds & params & 7-class & loud & difference [$95\%$ CI] & at cap \\
\midrule
recurrent, retrained & 2 & 0.10\,M & 0.254 & 0.432 & reference & no \\
recurrent, main run & 3 & 0.10\,M & 0.254 & 0.427 & -0.006 $[-0.016,+0.005]$ & no \\
recurrent, three blocks & 2 & 0.17\,M & 0.253 & 0.425 & -0.008 $[-0.021,+0.007]$ & no \\
recurrent, 160 epochs & 2 & 0.10\,M & 0.253 & 0.418 & -0.014 $[-0.027,-0.002]$ & no \\
pooled, source parameters & 2 & 0.10\,M & 0.244 & 0.416 & -0.016 $[-0.038,+0.002]$ & yes \\
pooled, coordinate channel & 2 & 0.09\,M & 0.254 & 0.415 & -0.017 $[-0.036,+0.003]$ & yes \\
pooled, both side inputs & 2 & 0.10\,M & 0.242 & 0.414 & -0.018 $[-0.039,-0.001]$ & no \\
pooled, double width & 2 & 0.28\,M & 0.247 & 0.408 & -0.025 $[-0.045,-0.008]$ & yes \\
pooled, light regularization & 2 & 0.09\,M & 0.246 & 0.408 & -0.025 $[-0.048,-0.003]$ & no \\
pooled, flattened read-out & 2 & 1.12\,M & 0.251 & 0.406 & -0.026 $[-0.044,-0.008]$ & no \\
pooled, response scale & 2 & 0.10\,M & 0.241 & 0.400 & -0.033 $[-0.049,-0.016]$ & no \\
pooled, classification only & 2 & 0.09\,M & 0.242 & 0.399 & -0.033 $[-0.048,-0.019]$ & no \\
residual network & 2 & 2.24\,M & 0.245 & 0.398 & -0.034 $[-0.054,-0.014]$ & no \\
pooled, 200 epochs & 2 & 0.09\,M & 0.241 & 0.393 & -0.039 $[-0.058,-0.023]$ & no \\
\bottomrule
\end{tabular}
\caption{Network variants on aLIGO. Accuracies are means over the seeds. The difference column gives the loud accuracy relative to the retrained recurrent network, with its paired source-bootstrap $95\%$ interval. The last column states whether any seed reached its best validation loss within one epoch of the maximum number of epochs.}
\label{tab:ablation}
\end{table}

Table~\ref{tab:ablation} compares the variants with the retrained recurrent network. None of them reaches a higher loud accuracy, and all remain below the residual bank of table~\ref{tab:headline}. The recurrent versions come closest, and their differences from the retrained network are small. Most pooled variants are lower by about two to four percentage points, and for most of them the bootstrap interval excludes zero. Adding the source parameters or the frequency position brings the pooled network closer to the recurrent one, but not above it.

The advantage of the recurrent network can be understood from the way the two read-outs summarize the band. The convolution blocks turn the response function into a sequence of local features along frequency. The six families differ mainly in where along the band their deviation is large, since the phase correction of family $b$ varies as $f^{b/3}$. For negative exponents it is largest at low frequencies, while for positive exponents it grows towards the end of the inspiral. The recurrent read-out goes through the features in the order of frequency and therefore keeps this information. Global average pooling, in contrast, replaces each feature by its average over the whole band, so that a feature at low frequencies and the same feature at high frequencies give the same result. The pooled variants are consistent with this picture. Those that come closest to the recurrent network receive additional information, such as the frequency position as an extra input channel, whereas making the pooled network wider or deeper does not help. The flattened read-out also keeps the position, but with about ten times more parameters it does not reach the recurrent network.

This comparison has two limitations. First, each variant was trained with only two seeds, and each interval in table~\ref{tab:ablation} treats one comparison on its own, so that small differences should not be over-interpreted. Second, most variants change several things at once. For instance, the classification-only variant differs from the recurrent network both in the read-out and in the absence of the regression loss, so that it does not tell whether the regression loss helps. The comparison therefore shows that the recurrent network is a good choice among those tested, but not which of its ingredients matters most.

\FloatBarrier
\section{Conclusions and outlook}
\label{sec:discussion}

In this work, we have extended the response-function approach of~\cite{hemmatyar_paper1_2026} from binary classification to a neural network with two outputs, which assigns a signal to GR or to one of six ppE phase families and estimates the magnitude of the coupling. We studied 262 catalog sources at the Advanced LIGO design sensitivity and 132 sources at the Hanford O3 sensitivity, in Gaussian noise and in recorded Hanford strain, with the intrinsic source parameters fixed at their injected values.

The network identifies the exponent of loud deviations in $43\%$ of the cases on aLIGO and in $33\%$ on Hanford. These values are modest, but they are close to those of the residual bank, which knows the parameters of each source and the exact residual of every family. The two classifiers make largely the same errors, and these errors follow the geometry of the families, since after the match neighboring exponents leave nearly parallel residuals. Moreover, thirteen network variants of different size and design all remain below the residual bank. The accuracy is therefore limited mainly by the similarity of the phase families rather than by the network, although our reference is approximate and does not provide a strict bound.

Coarser questions are answered much more reliably. A loud deviation is flagged in nearly every case, while about $1\%$ of the GR signals are flagged, and the sign of its exponent is identified in about $80\%$ of the loud aLIGO samples. The difficulty thus lies in the exact exponent rather than in the detection of a deviation or in its rough frequency dependence.

An initial analysis indicates that the identification improves considerably with the sensitivity of the detector. We applied the same construction to the same 262 sources with the Einstein Telescope ET-D sensitivity~\cite{hild_etd_2011,punturo_et_2010} and a lower cutoff of $2$~Hz, and with the Cosmic Explorer sensitivity~\cite{abbott_nextgen_2017,reitze_ce_2019} and a lower cutoff of $5$~Hz, each treated as a single detector. On loud samples, the network then identifies the exponent with accuracies of $0.63$ and $0.48$, compared with $0.43$ on aLIGO, and the residual bank shows the same trend. Two effects contribute to this improvement. First, the much larger signal-to-noise ratios allow the strength to reach its upper limit $s=20$ for essentially every source, so that a much larger fraction of the samples carry loud deviations, and the accuracy grows with the strength. Second, at equal residual strength the identification improves only for the Einstein Telescope, whose sensitivity extends to the lowest frequencies. Indeed, the phase families differ in their frequency dependence, and a longer observed inspiral separates them better, in agreement with the larger angles found in section~\ref{sec:geometry} for lighter sources and for the aLIGO curve compared with the Hanford curve. A detailed study of both detectors is work in progress.

The regression head does not estimate the coupling reliably, and an estimate conditioned on the predicted family is left for future work. In recorded Hanford noise, the network performs nearly as well as in Gaussian noise.

The main restriction of this work is the use of the true intrinsic parameters. Estimating them from the data will absorb part of each deviation and is expected to lower the accuracies. It is also the situation in which a trained network is most useful, since a comparison like the residual bank would then require a separate parameter estimation for every family. This analysis is the subject of a subsequent work. Further extensions include signals with higher modes and precession, networks of detectors, and architecture comparisons that change one ingredient at a time.

\acknowledgments
We acknowledge the funding from the European Research Council (ERC) under the European Union's Horizon 2020 research and innovation programme grant agreement No 801781. L.H. further acknowledges support from the Deutsche Forschungsgemeinschaft (DFG, German Research Foundation) under Germany's Excellence Strategy EXC 2181/1--390900948 (the Heidelberg STRUCTURES Excellence Cluster).

This research has made use of data or software obtained from the Gravitational Wave Open Science Center (gwosc.org), a service of the LIGO Scientific Collaboration, the Virgo Collaboration, and KAGRA. This material is based upon work supported by NSF's LIGO Laboratory which is a major facility fully funded by the National Science Foundation, as well as the Science and Technology Facilities Council (STFC) of the United Kingdom, the Max-Planck-Society (MPS), and the State of Niedersachsen/Germany for support of the construction of Advanced LIGO and construction and operation of the GEO600 detector. Additional support for Advanced LIGO was provided by the Australian Research Council. Virgo is funded, through the European Gravitational Observatory (EGO), by the French Centre National de Recherche Scientifique (CNRS), the Italian Istituto Nazionale di Fisica Nucleare (INFN) and the Dutch Nikhef, with contributions by institutions from Belgium, Germany, Greece, Hungary, Ireland, Japan, Monaco, Poland, Portugal, Spain. KAGRA is supported by Ministry of Education, Culture, Sports, Science and Technology (MEXT), Japan Society for the Promotion of Science (JSPS) in Japan; National Research Foundation (NRF) and Ministry of Science and ICT (MSIT) in Korea; Academia Sinica (AS) and National Science and Technology Council (NSTC) in Taiwan.

We acknowledge the use of Anthropic's Claude Fable 5.1 and Claude Opus 5.5, through Claude Code, for the design and generation of the plots, the generation of the tables, the refinement of the text and the checking of spelling and grammar.

\appendix
\section{Derivation of the mismatch response function}
\label{app:response}

In this appendix, we derive the mismatch response function of eq.~\eqref{eq:response}, following~\cite{hemmatyar_paper1_2026}, and its form in terms of the waveform residual $r=d-h_{\rm m}$. The data $d$, the spectrum $\Sn$ and the integration band are fixed throughout, and the intrinsic parameters of the template are those of the injected source.

The match of eq.~\eqref{eq:match} maximizes over the arrival time and, through the absolute value, over the phase of the template. At the maximizing time, we choose the phase of $h$ such that its complex overlap with the data is real and positive, and we call this the aligned template, for which $\mathcal{M}=\ip{h}{d}/(\norm{h}\norm{d})$. At the maximum, the derivatives with respect to time and phase vanish, so that small changes of the optimal time and phase do not change the match at first order. We can therefore vary the template with its alignment held fixed.

Let $\widehat h=h/\norm{h}$ and $\widehat d=d/\norm{d}$. Under a small change $\Delta h$ of the template, its norm and its normalized form change by
\begin{equation}
\delta\norm{h}=\frac{\ip{h}{\Delta h}}{\norm{h}},\qquad
\delta\widehat h=\frac{\Delta h}{\norm{h}}
-\frac{h\,\ip{h}{\Delta h}}{\norm{h}^3}.
\label{eq:response_norm_variation}
\end{equation}
Since the data are fixed, $\delta\bar{\mathcal{M}}=-\ip{\delta\widehat h}{\widehat d}$, and inserting eq.~\eqref{eq:response_norm_variation} gives eq.~\eqref{eq:dmismatch}. Written as a frequency integral, it reads
\begin{equation}
\delta\bar{\mathcal{M}}
=\operatorname{Re}\int\frac{4\,\Delta h(f)}{\norm{h}\Sn(f)}
\left[-\frac{d^*(f)}{\norm{d}}
+\mathcal{M}\frac{h^*(f)}{\norm{h}}\right]df.
\label{eq:response_general_variation}
\end{equation}
For a small phase change $h\to h\exp[i\epsilon\xi(f)]$, with a real profile $\xi(f)$ and a small real parameter $\epsilon$, we have $\Delta h=i\epsilon h\xi$, and eq.~\eqref{eq:response_general_variation} becomes
\begin{equation}
\delta\bar{\mathcal{M}}=\epsilon\,\operatorname{Re}\int R(f)\xi(f)\,df,\qquad
R(f)=\frac{4ih(f)}{\norm{h}\Sn(f)}
\left[-\frac{d^*(f)}{\norm{d}}
+\mathcal{M}\frac{h^*(f)}{\norm{h}}\right].
\label{eq:response_phase_derivation}
\end{equation}
This is the mismatch response function of eq.~\eqref{eq:response}. Since a phase change preserves the norm of the template, the term proportional to $\mathcal{M}$ contributes only to the imaginary part of $R$. Our phase correction $\dpsi=\beta\phi_b$ already contains $\beta$, so that no separate factor $\beta$ appears in eq.~\eqref{eq:response_def}, unlike in~\cite{hemmatyar_paper1_2026}.

We now evaluate the response function at the matched template $h_{\rm m}$ of eq.~\eqref{eq:amplitude}. With $B=\langle h_{t_0},d\rangle$ at the maximizing time, the squared residual norm for an arbitrary complex coefficient $a$ is
\begin{equation}
\norm{d-a h_{t_0}}^2
=\norm{d}^2-\frac{|B|^2}{\norm{h}^2}
+\norm{h}^2\left|a-\frac{B}{\norm{h}^2}\right|^2,
\label{eq:response_least_squares}
\end{equation}
where $\norm{h_{t_0}}=\norm{h}$. It is smallest for $a=B/\norm{h}^2$, which fixes both the amplitude and the phase. For $h_{\rm m}=a h_{t_0}$ and $r=d-h_{\rm m}$, we therefore obtain
\begin{equation}
\langle h_{\rm m},d\rangle=\norm{h_{\rm m}}^2,\qquad
\langle h_{\rm m},r\rangle=0,\qquad
\mathcal{M}=\frac{|B|}{\norm{h}\norm{d}}
=\frac{\norm{h_{\rm m}}}{\norm{d}}.
\label{eq:response_matched_identities}
\end{equation}
The last equality holds only for this least-squares amplitude. With eq.~\eqref{eq:response_matched_identities}, the bracket in eq.~\eqref{eq:response_phase_derivation} becomes
\begin{equation}
-\frac{d^*(f)}{\norm{d}}+\mathcal{M}\frac{h_{\rm m}^*(f)}{\norm{h_{\rm m}}}
=-\frac{d^*(f)-h_{\rm m}^*(f)}{\norm{d}}
=-\frac{r^*(f)}{\norm{d}}.
\label{eq:response_residual_bracket}
\end{equation}
Consequently, the response function of each sample is
\begin{equation}
R(f)=-\frac{4i h_{\rm m}(f)r^*(f)}
{\norm{h_{\rm m}}\norm{d}\Sn(f)},
\label{eq:response_residual_derivation}
\end{equation}
which is eq.~\eqref{eq:response_used}.

\clearpage
\bibliographystyle{JHEP}
\bibliography{biblio}
\end{document}